# ADAIS: Automatic Derivation of Anisotropic Ideal Strength via high-throughput first-principles computations

S. H. Zhang[1, 2], Z. H. Fu[1, 2] and R. F. Zhang[1, 2, *]

[1]*School of Materials Science and Engineering, Beihang University, Beijing 100191, P. R. China*

[2]*Center for Integrated Computational Engineering (International Research Institute for Multidisciplinary Science) and Key Laboratory of High-Temperature Structural Materials & Coatings Technology (Ministry of Industry and Information Technology), Beihang University, Beijing 100191, P. R. China*

\* Corresponding author: zrf@buaa.edu.cn (R.F. Zhang)

**Abstract**

Anisotropic ideal strength is a fundamental and important plasticity parameter in scaling the intrinsic strength of strong crystalline materials, and is a potential descriptor in searching and designing novel hard/superhard materials. However, to the best of our knowledge, an automatic derivation of anisotropic ideal strength has not been implemented in any open-source code available so far. In this paper, we present our developed ADAIS code, an automatic derivation of anisotropic ideal strength via high-throughput first-principles computations for both three-dimensional and two-dimensional crystalline materials with any symmetry, as well as for an ideal interface model. Several fundamental mechanical quantities can be automatically derived, including ideal tensile and shear strengths through affine deformation, universal binding energy and generalized stacking fault energy, as well as the ideal cleavage and slide stresses through alias deformation. The implementation of this code has been comprehensively demonstrated and critically validated by a lot of evaluations and tests of various crystalline materials with different symmetry, indicating that our code could provide a high-efficiency solution to quantify the strength of strong solids.

**Program summary**

*Program title:* ADAIS

*Licensing provisions:* BSD 3-Clause

*Programming language:* Fortran90

*Nature of problem:* A scheme adapted to high-throughput first-principles computations is much necessary to automatically determine the anisotropic ideal strength along any crystallographic orientation under affine and alias deformations for crystalline solids with any symmetry.

*Solution method:* To derive the ideal strength of a crystalline solid along a specific crystallographic orientation, a projection and/or redefinition of a lattice are firstly performed. The affine (alias) deformation is then applied to the projected (redefined) lattice, and a relaxation is done on the distorted structure by the simply modified VASP code optimizer with certain cell constraints. Afterwards, the resultant total energies and stresses *vs* strain are generated accordingly. Finally, the anisotropic ideal strengths are derived from the calculated stress/energy-strain relationship, and meanwhile the variation of bond length as a function of strain is revealed to illustrate the failure mode. Whenever an unexpected error happens, a message will be printed for further evaluation.

# 1. Introduction

The design of novel strong solids, e.g. ultraincompressible and hard materials, has recently attracted tremendous interest because of their fundamental importance in material science/physics and also their technological applications such as cutting and polishing tools in machinery industry, and drilling bits in milling and petrochemical industry [1-3]. Because of its high efficiency and low cost, high-throughput (HT) first-principles computations are becoming one of the most promising and impressive solutions to search the novel strong solids, most of which adopt the energy [4-6], elastic moduli [7] or empirical rules [8, 9] as descriptors. However, these descriptors do not always guarantee a high mechanical strength, due to their indirect correlation to the crystal plasticity, *i.e.* the irreversible mechanical response to external loadings [2]. To fulfill such necessity accordingly, the first-principles derivation of ideal strength provides a promising solution to quantitatively describe the plasticity-related behavior of a crystal under various deformations, which can also be adapted for the HT schemes.

Actually, ideal strength is an inherent mechanical property, which corresponds to the critical stress where a defect-free crystal becomes unstable and undergoes spontaneous structural transformation [10]. It provides insights into how the crystal is strongly bonded and where the lattice or electronic instability appears at large stain [11-13]. Furthermore, the ideal strength is closely related to the properties of dislocation mobility, e.g. the dislocation width and the dislocation nucleation [14]. From some specified well-defined experiments, e.g., nanoindentation [15] and nanostructured materials [16, 17], the ideal strength can be experimentally determined, showing excellent agreements with theoretical value, such as the experimental data of 13.4 GPa [16] *vs* theoretical value of 14.2 GPa [18] for the ideal tensile strength of Fe, and the experimental maximum flow stress of ~4.5 GPa [17] *vs* calculated pure shear strength of 2.8-4.9 GPa [19] for Ti. In the case of two-dimensional (2D) materials, the experimentally measured tensile strength of graphene monolayer is 130±10 GPa [20], being comparable to the theoretical value of 107-121 GPa [21]. The maximum strength for $MoS_2$ is experimentally obtained to be 15±3 N/m [22], which is consistent with the theoretical biaxial tensile strength of 15.2 N/m [23]. Nevertheless, it must be noted that the defects or flaws generally limit the strength of a real crystal far below its ideal value, and thus one may increase the strength by hindering or removing the crystal defects to approach the ideal strength.

Since initially explored by Esposito *et al.* [24] and Paxton *et al.* [25], the ideal strength under affine deformation has been widely used to explore the intrinsic mechanical response of crystalline solid and ideal interface system. For instance, the electronic instabilities of various transition metal borides, carbides and nitrides at large shear strain limit their achievable strength albeit they exhibit large elastic moduli, e.g., $OsB_2$ [26], $ReB_2$ [13], $WB_3$ [12], $CrB_4$ [11], $ZrB_{12}$ [27], WC [28] and $PtN_2$ [29]. Another representative application of ideal strength under affine deformation is to illustrate the stress-facilitated structural transformation, such as the B3 (zincblende) to B1 (rocksalt) and B4 (wurtzite) to B1 transformation appearing generally in the II-VI and III-V semiconductors [30, 31]. In regards to the interface systems, Zhang *et al.* [32, 33] systematically explored the decohesion and shear strength of superhard nc-$TiN/SiN_x$ heterostructures, and the Friedel oscillation appearing at interfaces was found to limit the strengths of superhard nanocomposites and heterostructures.

Besides of ideal strength under affine deformation, the ideal cleavage/slide stress across slip plane under alias deformation provides a more realistic description on the localized mechanical response to external loadings. The universal binding energy $\chi$ is originally estimated by Rose *et al.* [34] to characterize the bonding between crystallographic planes as cleaving along a given direction, while the ideal cleavage stress $\sigma_{ic}$, defined as $\sigma_{ic}=\max\{\nabla\chi\}$, can quantify the critical tensile stress to open two neighboring planes. In the case of alias shear deformation, the ideal slide stress $\sigma_{is}=\max\{-\nabla\gamma\}$ denotes the maximum stress for the sliding between two neighboring atomic planes, where the generalized stacking fault energy (GSFE) $\gamma$ proposed by Vitek *et al.* [35] provides a quantification in depicting the energy variation when one part of crystal is rigidly sliding with respect to the other part along a given crystallographic plane. In addition, by introducing the GSFE into Peierls-Nabarro (P-N) model [36, 37], the Peierls stress could be derived, which corresponds to dislocation facilitated plastic deformation and is defined as the minimum stress for irreversible movement of dislocation with a Burgers vector at 0 K [38].

Although anisotropic ideal strength has been extensively employed to scale the intrinsic strength of strong crystalline solids, and is potential to be a good descriptor in searching of novel hard/superhard materials via HT first-principles computations, to the best of our knowledge, an automatic derivation of anisotropic ideal strength has not been implemented in any open-source code available so far. Therefore, we here present our developed ADAIS code, an automatic derivation of anisotropic ideal strength via high-throughput first-principles computations, to meet the demand to scale and in-depth understand the intrinsic strength for both three-dimensional (3D) and 2D crystalline materials, as well as for an ideal interface model. In Section 2, we shall firstly give an overview on the theoretical methods of ideal tensile and shear strengths via affine deformation, and ideal cleavage and slide stresses via alias deformation. Then, Section 3 depicts the automated scheme and workflow of ADAIS code, and afterwards, several comprehensive evaluations and tests will be provided in Section 4: the implementations of ADAIS code for the ideal tensile and shear strengths by affine deformation are given in Section 4.1, those for the binding energy and GSFE by alias deformation are in Section 4.2, and the ideal strength under high pressure in Section 4.3. In the last Section 5, a brief summary is given with a few remarks on the further development of ADAIS code.

## 2. Overview of theoretical models and methods

### 2.1 Ideal tensile and shear strengths by affine deformation

In general, the ideal tensile and shear strengths could be calculated via affine deformation, *i.e.*, all the layers of the crystal cell are displaced uniformly along the tensile or shear direction, as shown in Fig. 1. It is seen that during affine deformation, only the lattice vectors need to be changed, while the fractional coordinates of the atomic positions keep unchanged [39]. Hence, the affine deformation can be imposed by transforming the initial lattice vector matrix $R^{ini}$ to the deformed lattice vector matrix $R^{def}$ through operating the strain matrix $\varepsilon$ as following:

$$R^{def} = R^{ini} \cdot (I + \varepsilon), \tag{1}$$

where $I$ stands for a 3×3 identity matrix. For an affine tensile deformation, the strain matrix $\varepsilon$ is expressed as

$$\varepsilon_{tensile} = \begin{bmatrix} \varepsilon_1 & 0 & 0 \\ 0 & \varepsilon_2 & 0 \\ 0 & 0 & \varepsilon_3 \end{bmatrix}. \tag{2}$$

For affine shear deformation, two different shear modes, *i.e.*, pure and simple shear [40], are mostly utilized in the previous literatures. As shown in Fig. 1, for pure shear, a symmetrical displacement with equal magnitude is applied between two orthogonal Cartesian axes, while in case of simple shear, a parallel displacement is introduced for each atomic plane along one Cartesian axis [40]. The corresponding deformed matrices are expressed as

$$\varepsilon_{pure} = \begin{bmatrix} 0 & \tfrac{1}{2}\varepsilon_6 & \tfrac{1}{2}\varepsilon_5 \\ \tfrac{1}{2}\varepsilon_6 & 0 & \tfrac{1}{2}\varepsilon_4 \\ \tfrac{1}{2}\varepsilon_5 & \tfrac{1}{2}\varepsilon_4 & 0 \end{bmatrix} \text{ and } \varepsilon_{simple} = \begin{bmatrix} 0 & \varepsilon_6 & \varepsilon_5 \\ 0 & 0 & \varepsilon_4 \\ 0 & 0 & 0 \end{bmatrix}, \tag{3}$$

respectively, where $\varepsilon_1=\varepsilon_{xx}$, $\varepsilon_2=\varepsilon_{yy}$, $\varepsilon_3=\varepsilon_{zz}$, $\varepsilon_4=\varepsilon_{yz}+\varepsilon_{zy}$, $\varepsilon_5=\varepsilon_{zx}+\varepsilon_{xz}$, and $\varepsilon_6=\varepsilon_{xy}+\varepsilon_{yx}$ are in Voigt notation [41]. For 2D materials, the in-plane tensile strain matrix $\varepsilon_{2D}$ is defined as

$$\varepsilon_{2D}=\begin{bmatrix} \varepsilon_1 & 0 \\ 0 & \varepsilon_2 \end{bmatrix}. \qquad (4)$$

Uniaxial and biaxial tensile deformations for 2D materials are mostly used, and for uniaxial tensile deformation, $\varepsilon_1=0$ and $\varepsilon_2 \neq 0$, while for biaxial tensile deformation, $\varepsilon_1=\varepsilon_2 \neq 0$.

The ideal tensile strength normal to the weakest plane or the ideal shear strength along the easiest slip system is generally regarded as one plasticity-related quantity, differing from the elastic properties that represent a reversible mechanical response nearly at equilibrium. The anisotropic ideal tensile or shear strengths along specific crystallographic direction or slip system is firstly determined, and then the minimum ideal strength, that is mostly relevant to the fracture and plasticity in reality, can be obtained in comparison:

$$\sigma_{min}=\min\{\sigma_{[uvw]}\} \text{ and } \tau_{min}=\min\{\tau_{(hkl)[uvw]}\}, \qquad (5)$$

where [$uvw$] is the crystallographic direction and ($hkl$) is the slip plane. A higher value of $\sigma_{min}$ and $\tau_{min}$ indicates a higher resistance to the lattice instability by fracture or shear, indicative of a potentially higher hardness.

## 2.2 Ideal cleavage and slide stresses by alias deformation

In the preceding introduction of ideal tensile and shear strengths, an affine deformation is applied, indicating that all the atomic planes are involved in the homogeneous deformation. In contrast, the alias deformation involves only between two neighboring atomic planes for a crystal or an interface of composite, while the other layers or slabs remain their original relative positions (see Fig. 1). In order to distinguish from the aforementioned ideal strength by affine deformation, we define the generated stress from the alias deformation as **ideal cleavage stress** in tension **and ideal slide stress** in shear.

To calculate the ideal cleavage stress via alias tensile deformation, as shown in Fig. 1, two neighboring crystallographic planes are cleaved along a given direction with a separation distance, $d$. Then the total energy of the cleaved structure is calculated by means of first-principles method, and accordingly the universal binding energy $\chi(d)$ as a function of $d$ is obtained as

$$\chi(d) = \frac{E_C(d) - E_0}{A}, \quad (6)$$

and it can be further expressed as [34]

$$\chi(d) = \chi_C \left[ 1 - (1 + \frac{d}{d_0}) \exp(-\frac{d}{d_0}) \right], \quad (7)$$

where, $E_C$ ($E_0$) is the energy of the cleaved (perfect) structure, and $A$ is the area of the cleaved plane. $\chi_C \equiv \lim_{d \to \infty} \chi(d)$ is defined as the cleavage energy. The stress $\sigma$ can be calculated by $\sigma(d) = \nabla \chi(d)$, and $d_0$ is the critical spacing at which the stress reaches its maximum. The ideal cleavage stress $\sigma_{ic}$, defined as $\max\{\sigma(d)\}$, represents the critical tensile stress needed to cut the bonds between the given cleavage planes.

By means of alias shear deformation, one can derive the GSFE and ideal slide stress, as shown in Fig. 1. The GSFE is a critical energetic quantity that depicts the energy variation when one part of crystal is rigidly sliding with respect to the other part along a given crystallographic direction [42], and it can be expressed as a function of displacement $u$:

$$\gamma(u) = \frac{E_{SF}(u) - E_0}{A}, \quad (8)$$

where $E_{SF}$ ($E_0$) is the energy of the slipped (perfect) structure, and $A$ is the area of the slip plane. As demonstrated by Vitek [35], the restoring force introduced in the P-N model [36, 37] is simply the gradient of the GSFE $\gamma(u)$:

$$\tau(u) = -\nabla \gamma(u). \quad (9)$$

The maximum slope $\tau^{max} = \max\{\tau(u)\}$ namely the ideal slide stress $\tau_{is}$, can be identified as the theoretical shear strength for the rigid interplanar sliding of a crystal along a specific slip direction.

Furthermore, the cleavage energy $\chi_C$ is the energy barrier to separate two neighboring crystallographic planes (brittle fracture), while the unstable GSFE $\gamma_{US}$ is the maximum energy for the sliding between two neighboring atomic planes, and represents the lattice resistance to the emission of a dislocation near the crack tip (plastic deformation) [43, 44]. Therefore, a non-linear relationship between Vickers hardness $H_v$ and $\chi_C$ or $\gamma_{US}$ was proposed in Ref. [43] as

$$H_v \propto \chi_C^n \text{ and } H_v \propto \gamma_{US}^n. \tag{10}$$

However, the cleavage resistance of a given crystal direction or slide resistance of a specified slip system cannot be quantified simply by the cleavage energy or unstable GSFE, because the magnitude of alias tensile and shear vectors may also play important roles. Hence, the ideal cleavage or slide stress is more relevant to infer the cleavage or slide resistance, and a relationship to scale the hardness of a crystal may result in the following form:

$$H_v \propto \min\{\sigma_{ic[uvw]}, \tau_{is(hkl)[uvw]}\}. \tag{11}$$

## 3. Implementations and workflows

We next describe the automated scheme and workflow of the ADAIS code using affine and alias deformations to determine the anisotropic ideal strength, binding energy, GSFE and other plastic properties (see Table 1). An automated procedure with minimum input parameter is adapted to meet the demands of HT computation scheme. The workflow of ADAIS code is schematically shown in Fig. 2 and more details are discussed below:

*Specify 3D or 2D materials & Read structure data*

Type of 3D or 2D materials is firstly specified for the calculation of anisotropic ideal strength. After that, the information of lattice vectors and atomic positions is read from the input structure file with a format of standard POSCAR file. Note that the crystal structure to calculate anisotropic ideal strength needs to be fully relaxed for both lattice parameters and atomic positions before reading structure data. In addition, for 2D materials, a sufficiently large vacuum layer is necessary to eliminate the interactions between the atomic layer and its periodic images.

*Determine symmetry & Redefine to IEEE-format*

To determine the possible slip system based on symmetry, firstly, the space group of the crystal structure is analyzed via the SPGLIB code [45]. Then, accordingly the structure will be redefined to unit cell with IEEE-format defined in Ref. [46], as the anisotropic ideal strength depends on the choice of coordinate system and lattice vectors. Note that, the *c*-axis of 2D materials is defined to be perpendicular to the atomic layers.

*Specify deformation mode: affine or alias*

Two deformation modes, *i.e.*, affine and alias, are supported in ADAIS code. For affine deformation, pure and simple deformations of 3D materials, and uniaxial and biaxial tensile deformations for 2D materials are alternative to meet the different demand of 3D and 2D crystalline materials. To calculate the ideal tensile or shear strength along a given direction, a projection procedure is firstly performed in order to make tensile crystallographic direction being parallel to one Cartesian axis (e.g. *y*-axis) for tensile deformation, while in the case of shear deformation, let the shear plane being normal to one Cartesian axis (e.g. *x*-axis) and keep the shear direction along another one (e.g. *y*-axis).

Two methods are implemented in ADAIS code to project the structure automatically: crystallographic index method and rotation method. For the crystallographic index method, a reciprocal lattice vector [*hkl*]\*, which is perpendicular to the lattice plane (*hkl*), and a lattice vector [*uvw*] are necessary. *h*, *k*, *l*, *u*, *v* and *w* must satisfy the condition: $hu+kv+lw=0$, to ensure the lattice vector [*uvw*] lies on the (*hkl*) plane. Then the reciprocal lattice vector [*hkl*]* and lattice vector [*uvw*] will be projected to be parallel to *x* and *y*-axes, respectively, as shown in Fig. 3. For the rotation method (see Fig. 3), three input parameters *α*, *β* and *θ* correspond to the contra-rotating angles along *x*, *y* and *z*-axes, respectively. Then the projected lattice vector matrix $R^{proj}$ is calculated from the initial lattice vector matrix R as following:

$$R^{proj}=R\cdot\begin{pmatrix}1 & 0 & 0\\ 0 & \cos\alpha & \sin\alpha\\ 0 & -\sin\alpha & \cos\alpha\end{pmatrix}\cdot\begin{pmatrix}\cos\beta & 0 & -\sin\beta\\ 0 & 1 & 0\\ \sin\beta & 0 & \cos\beta\end{pmatrix}\cdot\begin{pmatrix}\cos\theta & \sin\theta & 0\\ -\sin\theta & \cos\theta & 0\\ 0 & 0 & 1\end{pmatrix}. \qquad (12)$$

For alias deformation, both alias tensile and shear deformations are supported in ADAIS code, as shown in Fig. 1. Before applying alias deformation, a new right-handed lattice vector for 3D and 2D crystalline materials is redefined so that the cleavage or slip plane (*hkl*) is normal to *z*-axis (see Fig. 3). Meanwhile, to avoid the interaction of the cleavage or slip plane with its periodic images, a supercell of the redefined new lattice will be also created, for which the component of lattice vector *c* along *z*-axis is larger than 15.0 Å.

The mode with inputting one supercell structure (e.g., twin cell, stacking fault cell and interface cell) by the user and without further projecting for affine deformation or redefining new lattice for alias deformation, is also supported in ADAIS.

*Specify whether to relax or not & Prepare input files*

In ADAIS code, both relaxation and non-relaxation modes are supposed. In case of relaxation mode for affine deformation, both lattice vectors and atomic positions will be relaxed by modifying the five relevant strain components while remaining the applied strain component unchanged. The relaxation procedure will be terminated when the five conjugate Hellmann-Feynman stresses reach negligible values. In order to implement such relaxation automatically, several slightly modified VASP [47] optimizer with cell constraints are provided in ADAIS code to meet different conditions. In the case of non-relaxation mode, both crystal lattice and atomic position remain unrelaxed. In addition, it is also obtained to meet the condition of uniaxial strain that only the atomic coordinates are relaxed until the forces imposed on the atoms reach negligible values.

For the alias shear deformation, to get the entire GSFE profile including unstable GSFE, only the atomic movement normal to the slip plane is permitted in the case of relaxation mode. For the relaxation mode of alias tensile deformation, only the positions of the atoms, whose distance away from the cleavage plane is ≤4 Å, are permitted to relax. In addition, the non-relaxation mode for alias deformation is also implemented in ADAIS code to meet the condition of rigid cleavage and slide.

After that, the input files of VASP [47] are created for first-principles calculation. According to the specified relaxation mode, the corresponding INCAR file is created. To calculate the ideal strength under high pressure, one may input the pressure value as the INCAR file is created. In ADAIS code, two automatic methods to specify the KPOINTS file are obtained: the *k*-points per reciprocal atom (KPPRA) and the smallest allowed spacing between *k*-points (KSPACING). For more information of these two methods, one may refer the Ref. [46].

*Apply distortion to structure & Calculate total energy and stress*

For affine deformation, to get continuous strain path, the relaxed structure at previous strain step is used for the current strain step. Therefore, as calculating ideal strengths via affine deformation, one distortion is applied each time. Then, the total energy and stress of the distorted structure are calculated via first-principles code, and then are retrieved from the generated OUTCAR files. In addition, in case of relaxation mode, a filter of whether the structure relaxation is convergent or not, *i.e.*, the conjugate Hellmann-Feynman stresses reach negligible values, is also included in ADAIS code. If convergent, next strain step is processed using the relaxed structure at the current strain step and if not, a new density functional theory (DFT) calculation will be performed using the non-convergent structure. This procedure will be completed until the applied strain reaches to the maximum strain value specified by the user. Finally, the results of total energy and stress as a function of applied strain value are output to the file RDAIS.

Different from that of affine deformation, for alias deformation, a set of distortion values is applied to the crystal structure, and a series of distorted structure files are generated. Then, the DFT calculation will be performed. Finally, the total energy and stress of this series of distorted structures will be retrieved from the generated OUTCAR files, and the results of the binding energy $\chi(d)$ or GSFE $\gamma(u)$ will be output to the file RDAIS.

*Bond length analysis & Write the output files*

A filter is used in ADAIS code to justify whether the calculation is successful or not. If successful, the variation of bond length as a function of strain value under affine deformation, is revealed to determine the failure mode, and the results will be output to the file RBOND. If not, an error message will be printed for evaluation.

An example of the RBOND file for bct-C4 under affine pure shear deformation along the weakest (101)[10-1] slip system, is shown in Fig. 4a. It is seen that the instability of material under strain is always corresponding to a sharp variation of bond length. For bct-C4, the bonds between atoms 005 and 007 are broken at the stain value of 0.28, and meanwhile a process of graphitization happens with a C-C bond length of 1.413 Å (see Fig. 4b).

## 4. Evaluations and discussions

In this section, the implementation of ADAIS code was evaluated and tested on a broad class of materials. Our first-principles DFT calculations were performed using the VASP code [47] with the same calculation parameters as in our previous publication [46], except the calculation of GSFE for graphene bilayer. To compare the results calculated via

ADAIS code with the previous results in the literature [48] more realistically, the calculation condition in the literature [48], *i.e.*, the PBE functional with an energy cutoff of 500 eV and 36×24×1 *k*-mesh grids, and the van der Waals density functional of DFT-D2, were employed to calculate the GSFE of graphene bilayer.

## 4.1 Ideal tensile and shear strengths by affine deformation

For 3D materials, Table 2 lists the calculated ideal tensile and shear strengths via affine pure deformation of a broad class of materials with different symmetry except triclinic system, together with the previous theoretical values [4, 7, 27, 49-53] for comparison. It is found that all calculated values by ADAIS code are in reasonable agreement with the previous theoretical values [4, 7, 27, 49-53], confirming the validity of ADAIS code for 3D materials. Also, Table 3 lists the calculated uniaxial and biaxial tensile strengths of 2D materials with hexagonal and rectangular systems, together with the previous theoretical values [23, 54] for comparison. It is shown that all the theoretical values for the ideal strengths of 2D materials via uniaxial and biaxial tensile deformations show a good agreement with the previous theoretical values [23, 54].

In addition, Fig. 5a shows the calculated stress-strain curves of diamond along the weakest tensile (solid) and shear (open) path via affine pure (solid line) and simple (dished line) shear deformations. It is indicated that the ideal tensile strength of diamond along [111] direction is 92.4 GPa, and the stress-strain curves with affine simple shear deformation of (111)[11-2] and (11-2)[111] slip systems distribute on the both side of the associated pure shear curve. In ADAIS code, the total energy and volume as a function of strain are also output for 3D materials, and the results of diamond are illustrated in Fig. 5b. It is found that the instability in stress-strain curve is always corresponding to the break in the energy-strain and volume-strain curves. Fig. 5c shows the ideal strength via affine pure shear (red line) and tensile (blue line) deformations along different directions on the (111) plane. It is revealed that the lowest shear strength (93.6 GPa) is indeed along the directions of 0°, 120° and 240° deviating from the [11-2] direction, due to the symmetry of diamond. While for ideal tensile strength, it is shown a hexagonal symmetry with minimum and maximum values of 100.2 and 126.0 GPa along the direction of 0° and 30° deviating from the [11-2] direction, respectively. Therefore, the value of ideal tensile and shear strengths is much dependent on the direction.

The calculated stress-strain curves of graphene via uniaxial tensile deformations along X and Y directions and biaxial tensile deformation are shown in Fig. 6a. It is seen that the ideal strengths of uniaxial X, uniaxial Y and biaxial tensile deformations are 40.8, 36.8 and 34.2 N/m, respectively, which agree well with the previous calculated theoretical values, *i.e.*, 38.0, 34.4 and 32.0 N/m [23]. The calculated uniaxial tensile strength along different directions on the (0001) plane of graphene is shown in Fig. 6b, indicating that graphene has a largest uniaxial tensile strength (40.8 N/m) along the armchair direction, and a lowest uniaxial tensile strength (36.8 N/m) along the zigzag direction.

## 4.2 Binding energy and GSFE by alias deformation

Table 4 illustrates the calculated cleavage energy and GSFE of various materials with different symmetry, together with the previous theoretical values [42, 43, 55-59] for comparison, including both the unrelaxed and relaxed results. It is seen that all calculated values by ADAIS code agree with the previous theoretical values [42, 43, 55-59], confirming the validity of ADAIS code for alias deformation.

Fig. 7a shows the calculated binding energy of diamond (111) plane via alias tensile deformation using ADAIS code. To get the relaxed results, only the positions of the atoms, for which the distance away from the cleavage plane is ≤4 Å, are relaxed. It is found that diamond (111) plane has cleavage energies of 16.0 and 12.7 J/m$^2$ without and with relaxation, respectively, indicating the obvious influence of relaxation on the cleavage energy. Our calculated results show a reasonable agreement with the previous relaxed values of 14.2 J/m$^2$ [43]. The calculated GSFE profile of diamond (111)[1-10] slip system is also illustrated in Fig. 7b. To get the relaxed results, only the movement normal to the slip plane was permitted during atomic relaxation. It is found that diamond has the unstable GSFE $\gamma_{US}$ of 11.8 and 9.4 J/m$^2$ along (111)[1-10] slip system for the unrelaxed and relaxed results, respectively, indicating the necessity of relaxation for calculating the GSFE.

For 2D materials, Fig. 8 shows the calculated GSFE of graphene bilayer along the [10-10] direction, together with the relaxed interlayer spacing as a function of displacement. The previous theoretical values by Zhou *et al.* [48] are also illustrated in Fig. 8 for comparison. It is seen that an agreement is obtained for the GSFE and interlayer spacing of graphene bilayer. The equilibrium interlayer spacing of the stable structure AB for graphene bilayer is 3.26 Å, which agrees well with the previous theoretical value of 3.25 Å [48]. In addition, it is also found that the relaxed interlayer spacing curve mimics those of the corresponding GSFE curve.

Furthermore, the calculated γ-surface of TiC (111) plane is shown in Fig. 9a, which is a critical input parameter for the P-N model [36, 37]. It is found that the intrinsic stacking fault (ISF) is absent in TiC, which prevents the (111)[1-10] dislocation from splitting and increases the barrier of dislocation to motion [60]. In Fig. 9b, the calculated γ-surface of graphene bilayer is also shown, from which two minima at normalized displacements of 0 and 1/3 along (0001)[10-10] direction, are found within one period of displacement, and their corresponding structures are equivalent (*i.e.*, AB configuration). In addition, the preferred sliding direction of graphene bilayer is [10-10] with an energy barrier of 10.6 mJ/m$^2$, which is in reasonable agreement with the previous theoretical values of 12.2 mJ/m$^2$ [48].

## 4.3 Ideal strength under high pressure

Pressure/strain engineering has attracted increasing interest as an approach to improve the targeted properties of materials [42, 61, 62], therefore the calculation of ideal strength under high pressure is also implemented in ADAIS code. Fig. 10 shows the results of the ideal shear strength of diamond under different pressure along (111)[11-2] slip system, together with the previous theoretical values in Ref. [7] for comparison. It is seen that the ideal shear strength of diamond increases from 93.6 to 142.6 GPa as the pressure varies from 0 to 160 GPa, and an agreement with the previous theoretical values [7] is obtained indicating the successful implementation of ADAIS code for the calculation of ideal strength under high pressure. In addition, an increasing slope at near zero strain is also found as the pressure increases, corresponding to the profound pressure enhanced shear modulus [7] (see Fig. 10a).

## 5. Summary and perspective

In summary, our developed ADAIS code, an automatic scheme in deriving the anisotropic ideal strength adapted to HT first-principles computations is proposed for both 3D and 2D crystalline materials with any symmetry, as well as for an ideal interface model. A comprehensive evaluation on the implementation of this code has demonstrated its validity and efficiency in quantifying the strength of strong solids. In combination of our previously developed AELAS code [46], one may obtain sufficiently massive high quality data of mechanical properties for various crystals to meet the demands of the development of "Materials Genome Plan". We are currently advancing the ADAIS code to support the disturbed method [63] for ideal strength calculation, the calculation of the surface energy of any crystallographic plane and the analysis of phonon instability mode based on atomic displacement [63].

**Acknowledgements:**

This work is supported by the National Key Research and Development Program of China (Nos. 2017YFB0702100 and 2016YFC1102500), National Natural Science Foundation of China (NFSC) with No. 51672015, "111 Project" (No. B17002), National Thousand Young Talents Program of China, and Fundamental Research Funds for the Central Universities.

# Tables

Table 1. Properties of anisotropic ideal strength derived from ADAIS code for 3D and 2D crystalline materials.

| Models | Properties | Unit | Description |
|---|---|---|---|
| AFFINE | $E(\varepsilon)$ | eV | Total energy of distorted structure as a function of strain value $\varepsilon$ |
| | $V(\varepsilon)$ | Å$^3$ | Volume of distorted structure as a function of strain value $\varepsilon$ |
| | $\sigma_{[uvw]}$ | GPa | Ideal tensile strength under tensile deformation for 3D materials along crystallographic orientation [$uvw$] |
| | $\tau_{(hkl)[uvw]}$ | GPa | Ideal shear strength under pure/simple shear deformation for 3D materials along slip system ($hkl$)[$uvw$] |
| | $\sigma_{uniax}$ | GPa | Ideal tensile strength under uniaxial tensile deformation for 2D materials along specified orientation |
| | $\sigma_{biax}$ | GPa | Ideal tensile strength under biaxial tensile deformation for 2D materials |
| | $BL(\varepsilon)$ | Å | Variation of bond length as a function of strain value $\varepsilon$ |
| ALIAS | $\chi(d)$ | J/m$^2$ | Binding energy as a function of $d$ for specified cleavage plane |
| | $\sigma_{ic}$ | GPa | Ideal cleavage stress for specified cleavage plane; $\sigma_{ic}=\max(\nabla\chi(d))$ |
| | $\gamma(u)$ | J/m$^2$ | GSFE as a function of $u$ along specified slip system |
| | $\tau_{is}$ | GPa | Ideal slide stress along specified slip system; $\sigma_{is}=\max(-\nabla\gamma(u))$ |

Table 2. The calculated ideal tensile and shear strengths (in GPa) of various 3D materials with different symmetry via affine pure deformation, together with the previous theoretical values for comparison.

| Crystal system | Materials | Space Group | Pearson Symbol | Orientation | $\sigma_{min}$ | Slip system | $\tau_{min}$ | Ref. |
|---|---|---|---|---|---|---|---|---|
| Cubic | Diamond | *Fd-3m(#227)* | cF8 | [111] | 92.4 | (111)[11-2] | 93.6 | This work |
| | | | | | 90.7 | | 93.9 | Ref. [7] |
| | TiN | *Fm-3m(#225)* | cF8 | [100] | 32.2 | (111)[11-2] | 29.6 | This work |
| | | | | | 31.1 | | 28.8 | Ref. [49] |
| | ZrB$_{12}$ | *Fm-3m(#225)* | cF52 | [110] | 49.8 | (111)[11-2] | 35.4 | This work |
| | | | | | 48.9 | | 34.5 | Ref. [27] |
| Hexagonal | TiB$_2$ | *P6/mmm(#191)* | hP3 | [10-10] | 53.2 | (0001)[-12-10] | 49.7 | This work |
| | | | | | 54.1 | | 49.3 | Ref. [50] |
| | AlN | *P6$_3$mc(#186)* | hP4 | [-12-20] | 34.2 | (10-10)[-12-10] | 20.3 | This work |
| | | | | | 35 | | 20 | Ref. [49] |
| Trigonal | B$_6$O | *R-3m(#166)* | hR42 | [10-10] | 54.4 | (0001)[10-10] | 37.9 | This work |
| | | | | | 53.3 | | 38.0 | Ref. [51] |
| | R$_{13}$C$_2$ | *R-3m(#166)* | hR45 | [10-10] | 64.4 | (0001)[10-10] | 39.4 | This work |
| | | | | | 62.4 | | 39.4 | Ref. [51] |
| Tetragonal | bct-C$_4$ | *I4/mmm(#139)* | tI8 | [100] | 86.1 | (101)[10-1] | 87.4 | This work |
| | | | | | 85.8 | | 88.7 | Ref. [7] |
| | TiN$_2$ | *I4/mcm(#140)* | tI12 | [011] | 19.6 | (110)[1-11] | 15.5 | This work |
| | | | | | 19.2 | | 14.9 | Ref. [52] |
| Orthorhombic | IrB$_2$ | *Pmmn(#59)* | oP6 | | | (001)[100] | 8.0 | This work |
| | | | | | | | 7.9 | Ref. [4] |
| | OsB$_2$ | *Pmmn(#59)* | oP6 | [011] | 35.9 | (001)[100] | 9.1 | This work |
| | | | | | 37.6 | | 9.1 | Ref. [4] |
| | Z-Carbon | *Cmmm(#65)* | oC16 | [010] | 79.4 | (011)[01-1] | 85.7 | This work |
| | | | | | 79.5 | | 86.2 | Ref. [7] |
| | γ-B$_{28}$ | *Pnnm(#58)* | oP28 | [011] | 25.5 | (001)[010] | 23.7 | This work |
| | | | | | 25.3 | | 21.6 | Ref. [53] |
| Monoclinic | F-carbon | *P2/m(#10)* | mP8 | [100] | 73.5 | (001)[100] | 67.7 | This work |
| | | | | | 74.5 | | 68.2 | Ref. [7] |

Table 3. The calculated ideal strengths (in N/m) of various 2D materials with different symmetry via uniaxial tensile deformations along X and Y directions, and biaxial tensile deformation, together with the previous theoretical values for comparison.

| Crystal system | Materials | Uniaxial X | Uniaxial Y | Biaxial | Ref. |
|---|---|---|---|---|---|
| Hexagonal | Graphene | 40.8 | 36.8 | 34.2 | This work |
|  |  | 38.0 | 34.4 | 32.0 | Ref. [23] |
|  | $Ti_3C_2$ | 21.6 | 22.6 | 19.4 | This work |
|  |  | 21.6 | 22.5 | 19.6 | Ref. [54] |
|  | Silicene | 5.5 | 7.3 | 6.3 | This work |
|  |  | 5.3 | 6.9 | 6.1 | Ref. [23] |
|  | $MoS_2$ | 9.8 | 14.4 | 15.1 | This work |
|  |  | 9.7 | 14.3 | 15.2 | Ref. [23] |
|  | h-BN | 34.0 | 28.1 | 27.8 | This work |
|  |  | 33.8 | 28.1 | 27.7 | Ref. [23] |
| Rectangular | Borophene | 24.8 | 12.9 | 20.4 | This work |
|  |  | 22.3 | 12.8 | 23.9 | Ref. [23] |

Table 4. The calculated cleavage energy (in J/m$^2$) and GSFE (in J/m$^2$) of various 3D materials with different symmetry, together with the previous theoretical values for comparison.

| Materials | Orientation | Cleavage Energy | Slip system | GSFE | Ref. |
|---|---|---|---|---|---|
| Diamond | (111) | 12.7[a] | (111)[1-10] | 9.4[a] | This work |
|  |  | 14.2[a] |  | 9.1[a] | Ref. [43] |
| B$_6$O | (001) | 6.2[a] |  |  | This work |
|  |  | 6.6[a] |  |  | Ref. [55] |
| ReB$_2$ | (11-20)-S3 | 7.9[b] |  |  | This work |
|  |  | 8.9[b] |  |  | Ref. [56] |
| HfB$_2$ | (11-20) | 7.6[b] |  |  | This work |
|  |  | 7.7[b] |  |  | Ref. [56] |
| Mg |  |  | (0001)[10-10] | 0.034[a] | This work |
|  |  |  |  | 0.034[a] | Ref. [42] |
| Cu |  |  | (111)[11-2] | 0.043[a] | This work |
|  |  |  |  | 0.039[a] | Ref. [57] |
|  |  |  |  | 0.041-0.045[c] | Ref. [58] |
| SrTiO$_3$ |  |  | (110)[001] | 2.31[a] | This work |
|  |  |  |  | 3.05[a] | Ref. [59] |
|  |  |  | (110)[110] | 0.76[a] | This work |
|  |  |  |  | 1.02[a] | Ref. [59] |

[a] With structure relaxation;

[b] Without structure relaxation;

[c] Exp.

# Figures

Fig. 1. The general illustration of an ideal crystal for the calculation of ideal tensile and shear strengths via affine deformation, including affine tensile, pure and simple shear deformations. And the cleavage and glide processes for the crystallographic planes of an ideal crystal or interface by alias deformation, including alias tensile and shear deformations.

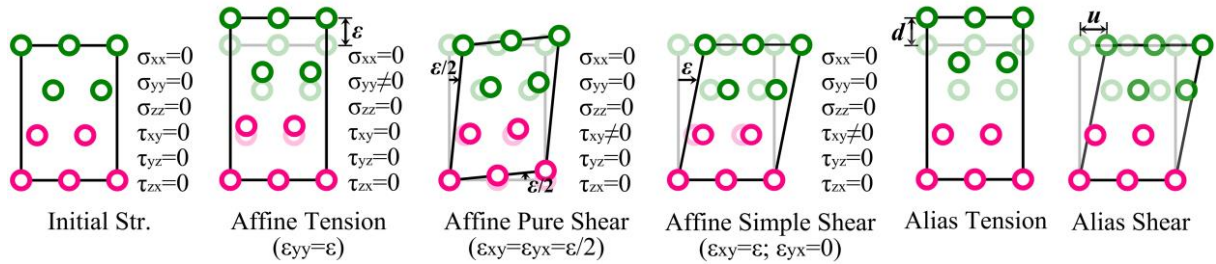

Fig. 2. Workflow of the algorithm used in ADAIS code.

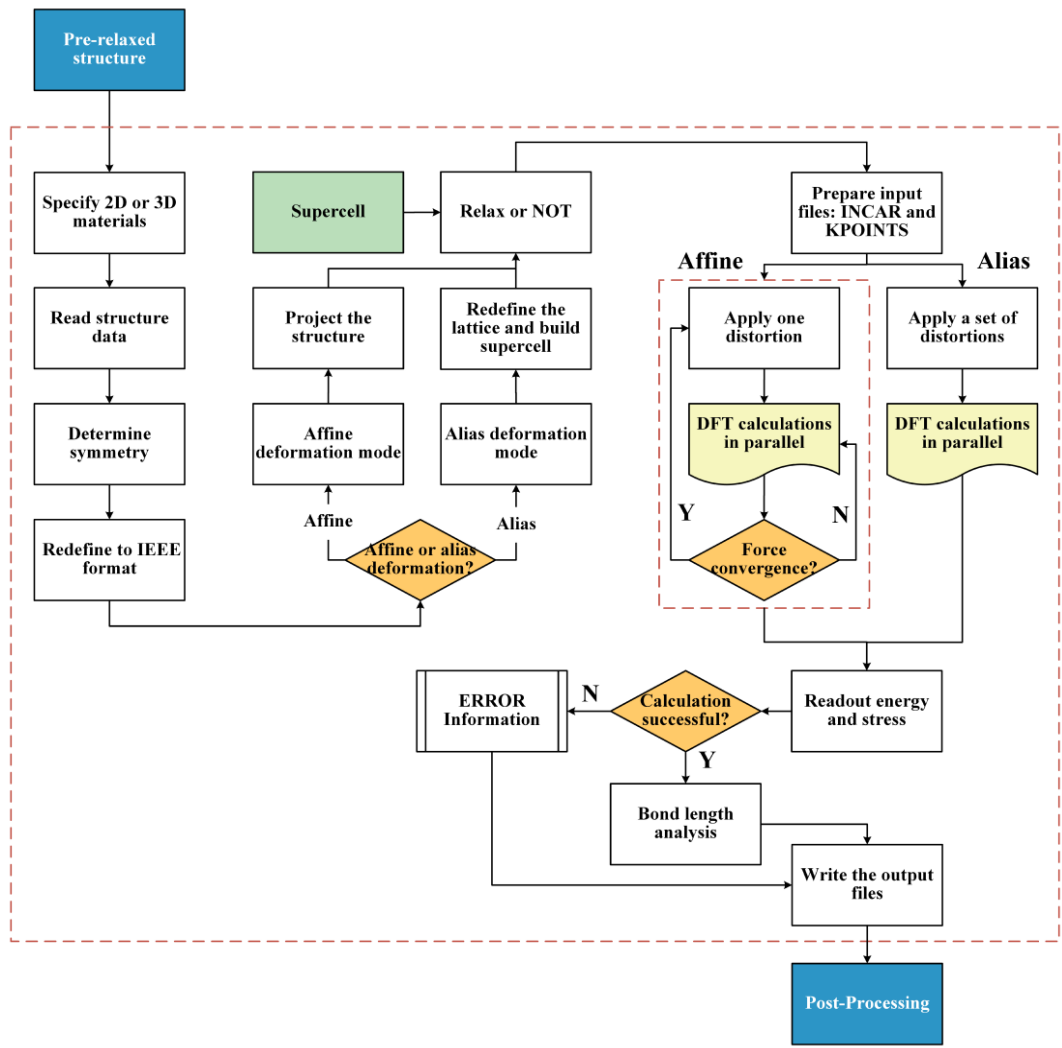

Fig. 3. The illustration of the projection, rotation and redefinition of a lattice used for affine and alias deformations: (i) Projection: the reciprocal-lattice vector [111]* and lattice vector [11-2] are projected to *x* and *y* axes, respectively; (ii) Rotation: the structure is rotated an angle of θ$_z$ along *z*-axis; (iii) Redefinition: the new *a*, *b* and *c* basis vectors are redefined as the lattice vectors of [111], [11-2] and [1-10], respectively.

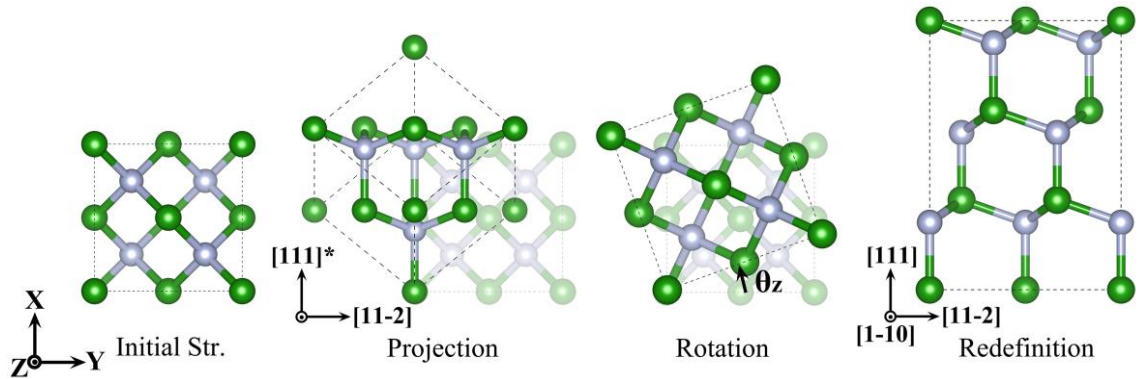

Fig. 4. (a) Example of the output file RBOND, which includes the variations of bond length as a function of strain value under affine deformation. (b) The variations of bond length as a function of strain, taking ATOM (005) as an example. (Inset) The corresponding atomic structures under affine pure shear strain of γ=0.0000 and γ=0.2800 (after instability).

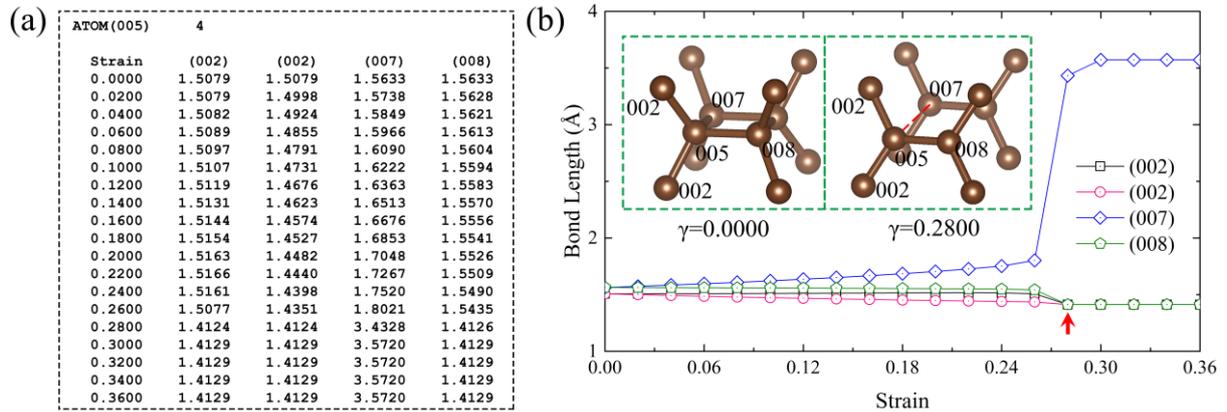

Fig. 5. (a) The calculated stress-strain curves of diamond along the weakest tensile (solid) and shear (open) paths via affine pure (solid line) and simple (dashed line) shear deformations. (b) The corresponding energy-strain and volume-strain curves of diamond. (c) The ideal strength via affine pure shear (red line) and tensile (blue line) deformations along different directions in the (111) plane.

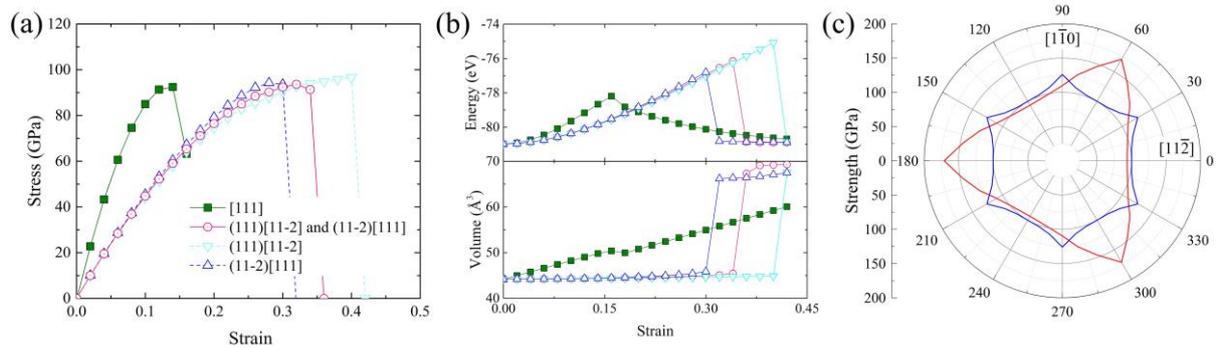

Fig. 6. (a) The calculated stress-strain curves of graphene via uniaxial tensile deformations along X and Y directions, and biaxial tensile deformation. (b) The anisotropic ideal uniaxial tensile strengths along different tensile directions on the (0001) plane of graphene. (Inset) Topological structure of graphene in top view, indicating that the X and Y directions correspond to the armchair and zigzag directions of graphene, respectively.

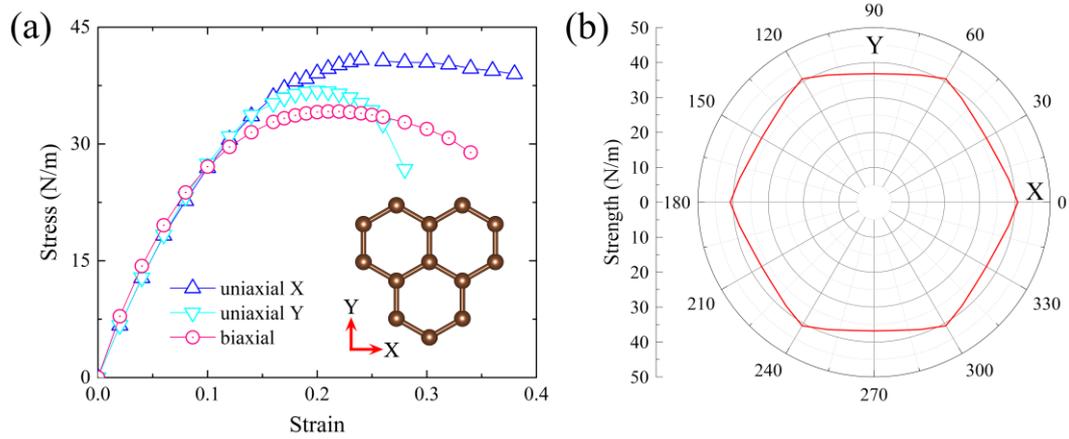

Fig. 7. (a) The calculated binding energy of diamond (111) plane without and with relaxation. For the relaxed calculation, only the positions of the atoms with a distance of ≤4 Å to the cleavage plane, were relaxed. (b) The calculated GSFE profile of diamond along (111)[1-10] slip system with unrelaxed and relaxed calculations. For the relaxed calculation, only the movement normal to the slip plane was permitted during atomic relaxation.

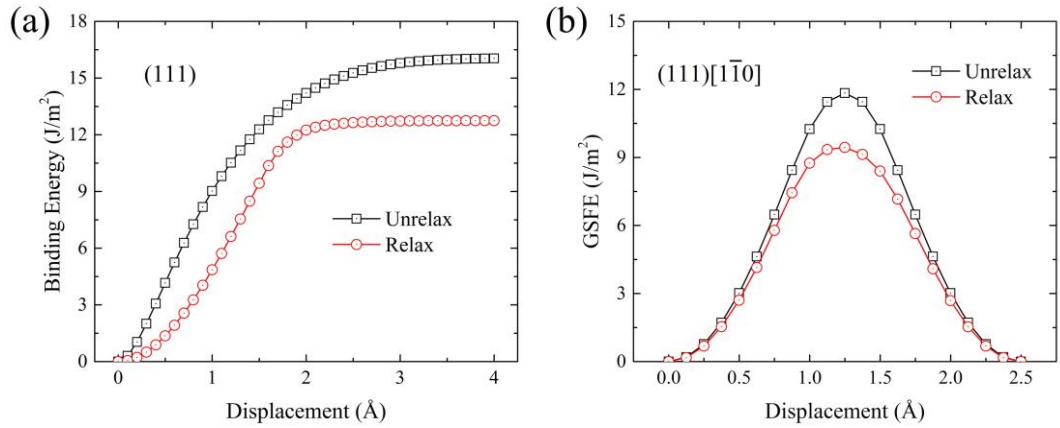

Fig. 8. The GSFE and the corresponding relaxed interlayer spacing *vs* displacement along the [10-10] direction for graphene bilayer, together with the previous theoretical values by Zhou *et al.* [48] for comparison. (Inset) The atomic structures of several high-symmetry configurations during alias shear deformation, *i.e.*, AB, SP and AA, are shown.

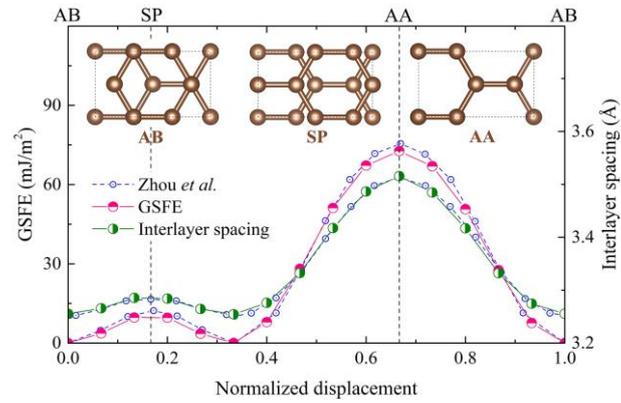

Fig. 9. The calculated γ-surface of (a) B1-TiC (111) plane and (b) graphene bilayer (0001) plane via ADAIS code.

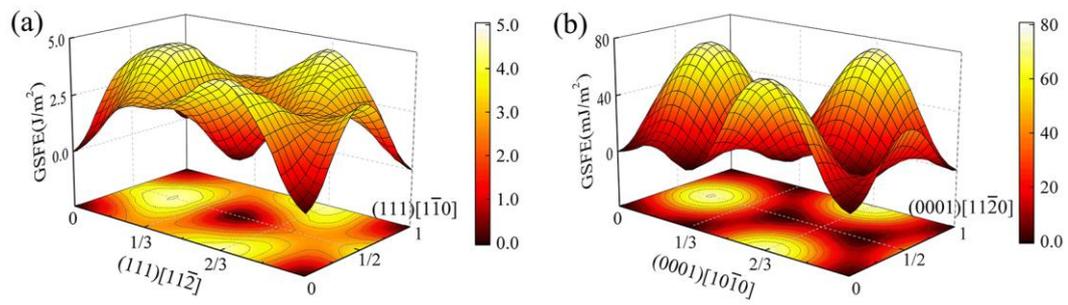

Fig. 10. (a) The calculated stress-strain carves of diamond via affine pure shear deformation along (111)[11-2] slip system under different pressures. (b) The variations of the calculated ideal strength of diamond along (111)[11-2] slip system under pressure ranging from 0 to 160 GPa, together with the previous theoretical values by Zhang *et al.* [7] for comparison.

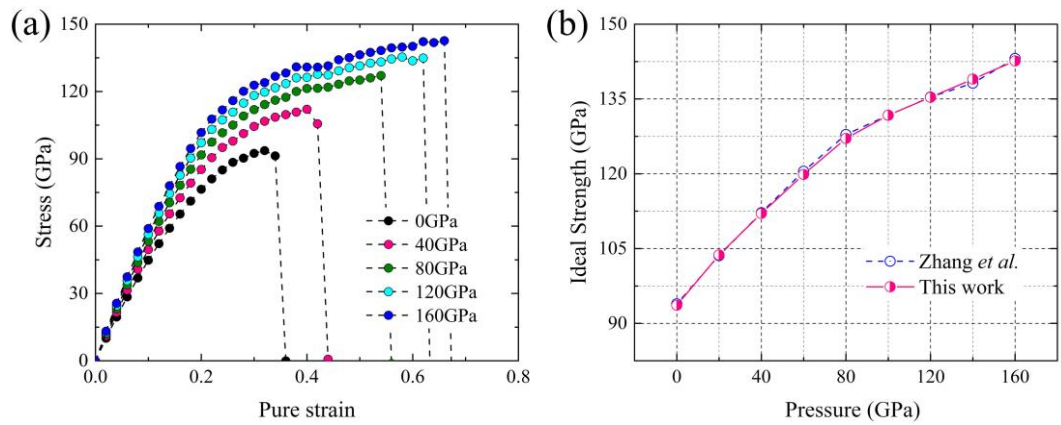